\documentclass{article}
\usepackage[preprint]{neurips_2025}

\usepackage[utf8]{inputenc} 
\usepackage[T1]{fontenc}    
\usepackage{hyperref}       
\usepackage{url}            
\usepackage{booktabs}       
\usepackage{amsfonts}       
\usepackage{nicefrac}       
\usepackage{microtype}      
\usepackage{xcolor}         
\usepackage{multirow}

\usepackage{caption}
\usepackage{hyperref}
\usepackage{appendix}

\usepackage[version=4]{mhchem} 

\usepackage{graphicx}

\title{Beyond Scaling: Chemical Intuition as Emergent Ability of Universal Machine Learning Interatomic Potentials}

\author{
  Shinnosuke Hattori \\
  Advanced Research Laboratory \\
  Sony Group Corporation \\
  Atsugi, Kanagawa, Japan \\
  \texttt{shinnosuke.hattori@sony.com} \
  \And
  Kohei Shimamura \\
  Department of Physics \\
  Kumamoto University \\
  Kumamoto, Japan
  \And
  Aiichiro Nakano \\
  Collaboratory for Advanced \\
  Computing and Simulation \\
  University of Southern California \\
  Los Angeles, CA, USA \\
  \And
  Rajiv K. Kalia \\
  Collaboratory for Advanced \\
  Computing and Simulation \\
  University of Southern California \\
  Los Angeles, CA, USA \\
  \And
  Priya Vashishta \\
  Collaboratory for Advanced \\
  Computing and Simulation \\
  University of Southern California \\
  Los Angeles, CA, USA \\
  \And
  Ken-ichi Nomura \\
  Collaboratory for Advanced \\
  Computing and Simulation \\
  University of Southern California \\
  Los Angeles, CA, USA \\
  \texttt{knomura@usc.edu}
}

\begin{document}

\maketitle

\begin{abstract}

Machine Learning Interatomic Potentials (MLIPs) have successfully demonstrated scaling behavior, i.e. the power-law improvement in training performance, however the emergence of novel capabilities at scale remains unexplored.
We have developed Edge-wise Emergent Decomposition (E3D) framework to investigate how an MLIP develops the ability to derive physically meaningful local representations of chemical bonds without explicit supervision. 
Employing an E(3)-equivariant network (Allegro) trained on molecular data (SPICE~2), we found that the trained MLIP spontaneously learned representations of bond dissociation energy (BDE) by decomposing the global potential energy landscape. The learned BDE values quantitatively agree with literature and its scalability are found to be robust across diverse training datasets, suggesting the presence of underlying representation that captures chemical reactions faithfully beyond given training information. Our E3D analysis utilizing Shannon's entropy reveals a close interplay between the decomposability of potential energy learning, scalability of learning, and emergent chemical reactivity, thus providing novel insights of scaling limitations and pathways toward more physically interpretable and predictive simulations.
\end{abstract}

\section{Introduction}

Machine Learning Interatomic Potentials (MLIPs) have revolutionized computational simulation by delivering near-quantum accuracy at substantially reduced computational cost~\cite{Behler-GeneralizedNeural-networkSurfaces-2007i,Bartok-GaussianApproximationElectrons-2010p,Batatia-FoundationModelChemistry-2023x}.
Recent progress has been driven primarily by scaling---increasing model size and dataset diversity---leading to predictable power-law improvements in accuracy~\cite{Merchant-ScalingDeepDiscovery-2023p,Kozinsky-ScalingLeadingSize-2023y}.
This scaling success has enabled novel software frameworks~\cite{Schutt-SchNetPack2Learning-2023u,Zeng-DeePMD-kitV2Models-2023m,Lysogorskiy-PerformantImplementationSilicon-2021j,Drautz-AtomicClusterPotentials-2019p,Batzner-E3-equivariantGraphPotentials-2022r,Batatia-MACEHigherFields-2022t, Kovacs-MACE-OFF23TransferableMolecules-2023f, Deng-CHGNetPretrainedModelling-2023g}, also large and diverse datasets including SPICE~\cite{Eastman-NutmegSPICELearning-2024y,Eastman-SPICEDatasetPotentials-2023j}, MPTrj~\cite{Deng-CHGNetPretrainedModelling-2023g}, Alexandria~\cite{Schmidt-Dataset175kFunctionals-2022j}, TM23~\cite{Owen-ComplexityMany-bodySet-2023q}, OMat24~\cite{Barroso-Luque-OpenMaterialsModels-2024w}, OMol25~\cite{Levine-OpenMoleculesModels-2025z}, and MatPES~\cite{Kaplan-FoundationalPotentialMaterials-2025p}.
These advances have produced increasingly generalizable MLIPs for applications ranging from battery materials~\cite{Ju2025-bz}, catalysts~\cite{Chanussot-OpenCatalystChallenges-2020b,Wander-CatTSunamiAcceleratingNetworks-2025x}, drug discovery~\cite{Gelzinyte-TransferableMachineMolecules-2024n,Hattori-RevisitingAspirinPotential-2024b}, and nanodevices~\cite{Kovacs-MACE-OFF23TransferableMolecules-2023f,Mortazavi-AtomisticModelingPotentials-2023i,Rodrigues-MachineLearningNanotubes-2025d,Shimamura-ThermalConductivityAllegro-2024k}.

While these remarkable applications using MLIPs have become available, a scientific question is: \textit{why and how do MLIPs generalize so well?} While current trends of MLIP development follow ``bigger and more diverse'' strategy~\cite{Merchant-ScalingDeepDiscovery-2023p, Kozinsky-ScalingLeadingSize-2023y,Zhang-ExploringFrontiersPotential-2024u}, it has become increasingly clear that the strategy faces challenges in the prediction of complex chemical reactions. ~\cite{Chanussot-OpenCatalystChallenges-2020b}.

Accurate prediction of chemical reactions is one of long-sought capabilities of MLIPs because the mechanistic understanding of reaction pathways and energy barriers may substantially accelerate the development and synthesis of novel materials, thereby reduce experimental costs. Great efforts have been made to date, such as large-scale data generation initiatives, e.g. the Open Catalyst Project~\cite{Chanussot-OpenCatalystChallenges-2020b,Tran-OpenCatalystElectrocatalysts-2023z,Wander-CatTSunamiAcceleratingNetworks-2025x}, and active learning for reaction modeling including ANI potential~\cite{Zhang-ExploringFrontiersPotential-2024u}. A recent development in ``foundation model'' has also demonstrated improved predictability of chemical reactions. These models have shown enhanced performance in chemical reaction prediction through training without specialized reaction-specific training data, often based on a hybrid dataset that combines organic and inorganic materials data~\cite{Nomura-Allegro-FMTowardsSimulations-2025u,Tomoya-TamingMulti-domainSimulations-2024e}. For example, Allegro-FM has demonstrated emergent capabilities for silicate systems, indicating that strategic data diversity —rather than simply increasing data volume for a specialized task— may enable models to learn more fundamental chemistry with broader transferability.

Despite these advancements, the lack of our understanding of how a model acquires chemical intuition during training remains as a critical challenge for scalable and predictable training accuracy across wide aspects of chemical reactions. A mechanistic understanding of MLIP learning is particularly crucial when distinct chemical properties, such as activation energy and reaction energy, are examined with increased training data volume and model capacity.

\begin{figure}[htbp]
    \centering
    \includegraphics[width=0.8\linewidth]{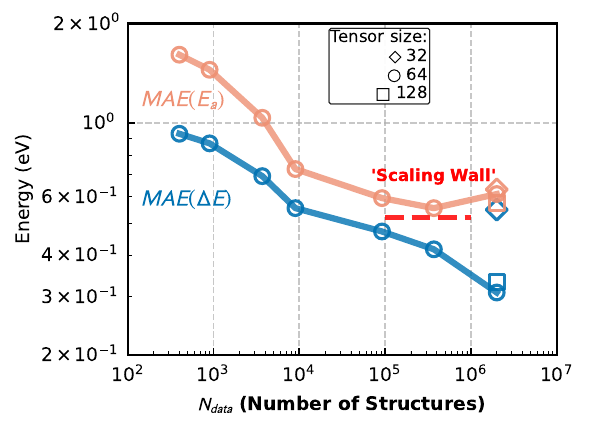}
    \caption{Scaling behavior of prediction errors for reaction energies ($\Delta E$) and activation barriers ($E_a$) on the Transition1x dataset~\cite{Schreiner-Transition1x-Potentials-2022b}. Mean Absolute Error (MAE) versus number of SPICE 2 training structures for Allegro models with tensor sizes 32 (diamonds), 64 (circles), and 128 (squares). While $\Delta E$ accuracy improves consistently, $E_a$ predictions hit a ``scaling wall.''}
    \label{fig:fig1}
\end{figure}

Figure~\ref{fig:fig1} shows a striking disparity in the scaling behavior between reaction energy and activation energy, i.e. the former continues to scale with respect to the dataset size while the latter shows a saturation point in MAE, measured with Allegro models trained on the Transition1x(T1x) dataset~\cite{Schreiner-Transition1x-Potentials-2022b}.

As shown in Figure~\ref{fig:fig1}, the accuracy of reaction energy ($\Delta E$) improves consistently with more training data across all model sizes, demonstrating the expected benefits of scaling. In stark contrast, activation barrier ($E_a$) accuracy plateaus after initial improvements, hitting a "scaling wall" where additional data provides little to no benefit. This disparity is observed consistently across larger parameter-space model, suggesting that the limitation is not merely a capacity issue but reflects a more fundamental challenge in how these models represent and learn reactive properties.

This observation of the scaling disparity has given rise to a number of questions and motivated this study; Why does the model successfully learn to predict $\Delta E$ while struggling with $E_a$? What differences in the underlying chemical processes or model representations account for this divergent scaling behavior? Is there a robust metric to capture the mechanisms behind these scaling disparities and help quantify chemical information that is internally learned by an MLIP?

To address these questions, we have developed Edge-wise Emergent Energy Decomposition (E3D) framework, with which the total energy is decomposed into symmetric ($D_{ij}$) and asymmetric ($A_{ij}$) representations of the bond energy. E3D allows us to categorize the chemical bond energy by their type (e.g., C-H, C-C bonds) thus ``Type-Decomposition'', quantify the strength of chemical bonds, and compare to Bond Dissociation Energies (BDEs) from literature. Taking advantage of the E(3)-equivariant Allegro architecture~\cite{Musaelian-LearningLocalDynamics-2023h} we employed in this study, we have analyzed distinct signatures of chemical bonds in the two-dimensional $D_{ij}$-$A_{ij}$ representational space.

A central finding of this study is that BDE learning occurs in a ''decomposable learning'' fashion --- a model acquires the ability to learn physically meaningful local representations for different chemical bond types during training without explicit supervision. BDE learning is also found to be remarkably robust, regardless of different training datasets, and correlates directly with various physical quantities. The E3D framework provides valuable insights on the scaling behavior, including explanations for the $E_a$ scaling wall, and offers actionable strategies to overcome existing limitations in MLIP development.

\section{Method}\label{sec:method}

\subsection{Background}\label{sec:method_model}
The potential energy 
$E_{\rm system}$ of a system composed of 
$N$ atoms can be described using a body-order expansion that includes terms up to the 
$N$-body interaction:
\begin{equation}
E_{\rm system} =
\sum_{i} E^{(1)}_{i}
+\sum_{ij} E^{(2)}_{ij} 
+\sum_{ijk} E^{(3)}_{ijk}
+ \cdot \cdot \cdot
+\ (N{\rm-body\ term}).
\label{eq:Eexpansion1}
\end{equation}

Since the computational cost of $N$-body interactions scales as $O(N^N)$, directly incorporating such terms into MLIPs is not practical.
Many of the current MLIPs allow the $E_{\rm system}$ to be expressed as a sum of single-atom energies $\tilde{E}^{(1)}_{i}$ that include many-body effects:
\begin{equation}
E_{\rm system} = \sum_{i} \tilde{E}^{(1)}_{i}.
\label{eq:Eexpansion_convMLIPs}
\end{equation}
To accomplish this, the existing MLIPs, such as Behler-Parrinello neural network potentials ~\cite{Behler-GeneralizedNeural-networkSurfaces-2007i}, describe many-body interactions by combining two-body and three-body descriptors through nonlinear functions, such as activation functions used in neural networks.
In particular,  Atomic Cluster Expansion (ACE) 
~\cite{Drautz-AtomicClusterPotentials-2019p}, which enables the construction of many-body descriptors in a computationally efficient manner, has significantly improved the accuracy of the representation in Eq.~\ref{eq:Eexpansion_convMLIPs}.

However, it is not necessary to represent the total energy with the single-atom energy $\tilde{E}^{(1)}_{i}$.
Here, the cohesive energy is defined as $E_{\rm coh} \equiv E_{\rm system} - \sum_{i} E^{(1)}_{i}$, and then Eq.~\ref{eq:Eexpansion1} becomes,
\begin{equation}
E_{\rm coh} = \sum_{ij} E^{(2)}_{ij} 
+\sum_{ijk} E^{(3)}_{ijk}
+ \cdot \cdot \cdot
+\ (N{\rm-body\ term}).
\label{eq:Eexpansion2}
\end{equation}
 
$E_{\rm coh}$ is preferred to $E_{\rm system}$ for the training target of MLIPs, because it eliminates dependencies on computational settings such as pseudopotentials in first-principles calculations, thereby providing an unbiased energy reference. 

Similar to that the representation from Eq.~\ref{eq:Eexpansion1} to Eq.~\ref{eq:Eexpansion_convMLIPs} was achieved, MLIPs should be able to expand $E_{\rm coh}$ in Eq.~\ref{eq:Eexpansion2} using two-body interactions $\tilde{E}^{(2)}_{ij}$ that include many-body effects, as follows:

\begin{equation}
E_{\rm coh} = \sum_{ij} \tilde{E}^{(2)}_{ij}.
\label{eq:Eexpansion3}
\end{equation}

Since $\tilde{E}^{(2)}_{ij}$ can be naturally interpreted as a quantitative measure of the attractive or repulsive interatomic interactions, 
Eq.~\ref{eq:Eexpansion3} serves as a fundamental concept for evaluating interatomic bond strengths using MLIPs themselves.

\subsection{Model and Energy Decomposition Formulation}\label{sec:method_model_allegro}
We employ the E(3)-equivariant Allegro architecture~\cite{Musaelian-LearningLocalDynamics-2023h} for our analysis due to its inherent energy decomposability shown in Eq.~\ref{eq:Eexpansion3}. We trained Allegro models with internal tensor representation sizes of 32, 64, and 128, spherical harmonics expansion $l_{\rm max} = 2$, and radial cutoff $r_{\rm cut} = 5.2$~\AA. For most analyses, we employed a tensor size of 64, prioritizing computational efficiency without compromising accuracy. This specific size proved optimal: a smaller tensor (size 32) resulted in lower accuracy due to its limited parameter count and thus reduced representational capacity, whereas a larger tensor (size 128) demonstrated comparable performance to size 64.

Allegro decomposes total system energy $E_{\rm system}$ into per-node (atom) energies $\varepsilon_i$ and per-edge energies $\varepsilon_{ij}$\cite{Musaelian-LearningLocalDynamics-2023h}:
\begin{equation}
\begin{split}
E_{\rm system} &= \sum_{i=1}^{N} (\sigma_{Z_i} \varepsilon_i + \mu_{Z_i}), \\
\varepsilon_i &= \sum_{j \in N(i)} \sigma_{Z_i Z_j}\varepsilon_{ij},
\end{split}
\label{eq:E_system}
\end{equation}
where $\sigma_{Z_i}$, $\sigma_{Z_i Z_j}$ are learnable per-species scale parameters, $\mu_{Z_i}$ are learnable shift parameters for atoms of species $Z_i$ and $Z_j$, $N$ is the total number of atoms, and $N(i)$ represents atom $i$'s local neighborhood.
The Allegro model, which is based on the ACE formalism~\cite{Drautz-AtomicClusterPotentials-2019p} and combined with their connection via nonlinear functions, enables the expression of $E_{\rm system}$ in terms of the per-edge energies $\varepsilon_{ij}$\cite{Musaelian-LearningLocalDynamics-2023h}.

Therefore, in order to implement an architecture that satisfies Eq.~\ref{eq:Eexpansion3} within the Allegro model, we standardize the normalization parameters by setting the shift parameters $\mu_{Z_i} = 0$ and scale factors $\sigma_{Z_i} = \sigma_{Z_i Z_j} = 1.0$ in Eq.~\ref{eq:E_system}.
Furthermore, instead of learning 
$E_{\rm system}$, the model is trained to predict 
$E_{\rm coh}$.
With these settings, Eq.~\ref{eq:E_system} becomes a direct sum of edge energies $\varepsilon_{ij}$:
\begin{equation}
E_{\rm coh} = \sum_{i=1}^{N} \sum_{j \in N(i)} \varepsilon_{ij},
\label{eq:Ei}
\end{equation}
which is equivalent to Eq.~\ref{eq:Eexpansion3}.

Importantly, $\varepsilon_{ij}$ and $\varepsilon_{ji}$ are generally not the same (i.e., $\varepsilon_{ij} \neq \varepsilon_{ji}$) reflecting  the local environments of ${i}$ and ${j}$ atoms independently. Thus, these energy values differ based on which atom index to be taken as the central atom. 
We define a set of metrics, i.e. the symmetric component $D_{ij}$ and asymmetric component $A_{ij}$, to capture the degree of the bond energy strength and its asymmetry:
\begin{equation}
\begin{split}
D_{ij} &= \varepsilon_{ij} + \varepsilon_{ji},\\
A_{ij} &= \varepsilon_{ij} - \varepsilon_{ji}.
\end{split}
\label{eq:DijAij}
\end{equation}
Here, $D_{ij}$ represents total bond energy between $i$th and $j$th atoms. 
The summation of $D_{ij}$ for all atomic pairs in the system, i.e., $\sum^N_i \sum_{j\in N(i),j > i} D_{ij}$ is equal to $E_{\rm coh}$.
The asymmetric component $A_{ij}$ quantifies energy imbalance arising from the difference in the local environment of the two atoms.  For example if $i$th and $j$th atoms are of the same element, $A_{ij}$ reflects the difference between the two local environments.

We trained the Allegro models modified with unscaled $\varepsilon_{ij}$ outputs using NequIP framework\cite{Batzner-E3-equivariantGraphPotentials-2022r}. Detailed training procedure and hyperparameters are provided in Appendix \ref{sec:ap_hyperparameter}, information about dataset are also provided in Appendix \ref{sec:ap_dataset}.

\subsection{E3D Analysis Protocol}\label{sec:analysis_protocol}

\paragraph{Bond Classification and BDE Comparison}
We assigned bond multiplicities (single, double, triple, aromatic) using General Amber Force Field (GAFF) atom types from \texttt{antechamber} in AmberTools~23\cite{Case-AmberTools-2023h} with manual validation for ambiguous cases. Reference BDE values from experimental measurements~\cite{Speight-LangesHandbookEdition-2016u} enable quantitative comparison with learned $D_{ij}$ distributions, providing a direct assessment of the physical meaningfulness of the learned representation.

In addition we define two key metrics, the BDE distribution shift $\Delta_{\text{BDE}}$ and the BDE distribution width $\sigma_{\text{BDE}}$, to enable a unified evaluation metric across different model training settings.
$\Delta_{\text{BDE}}$ and $\sigma_{\text{BDE}}$ are computed from a single distribution combining the BDE distributions of all bond types, each of which is shifted by their reference BDE value. See the inset in Figure~\ref{fig:fig3}.

\paragraph{2$D$ Map Analysis and Entropy Calculation}
To analyze correlations between the symmetric ($D_{ij}$) and asymmetric ($A_{ij}$) components of learned bond energies, we generated 2$D$ histograms of $D_{ij}$ versus $A_{ij}$. This analysis focused on \ce{C-C}, \ce{C-H}, \ce{C-O}, and \ce{C-N} s from T1x dataset structures, considering covalent bonds shorter than 2.2~\AA~to ensure chemical relevance. We constructed these histograms using uniform 0.3~eV energy bin widths for both axes over their observed ranges.

\paragraph{Information Entropy Evaluation}
Shannon entropy quantifies organization structure of representational spaces:
\begin{equation}
\begin{split}
H_{2D} &= - \sum_{k, l} p_{kl} \log p_{kl} \\
\end{split}
\label{eq:2Dentropy}
\end{equation}
where $p_{kl}$ is the normalized frequency in bin $(k,l)$ of the 2$D$ histogram. Lower entropy values indicate more organized, well-defined internal representations, suggesting that the model has developed structured chemical understanding. This metric provides a quantitative measure of how clearly the model distinguishes different chemical environments and s.

Table~\ref{tab:e3d_metrics} summarizes the metrics employed in the E3D framework to characterize learned representations and their relationship to physical quantities.

\begin{table}[htbp]
    \centering
    \caption{Key Metrics for E3D Analysis.}
    \label{tab:e3d_metrics}
    \begin{tabular}{p{0.3\textwidth}lp{0.4\textwidth}}
        \toprule
        Metric & Symbol & Description \\
        \midrule
        Symmetric Bond Energy & $D_{ij}$ & Total interaction energy: $\varepsilon_{ij} + \varepsilon_{ji}$ \\
        Asymmetric Bond Energy & $A_{ij}$ & Energy imbalance due to asymmetry: $\varepsilon_{ij} - \varepsilon_{ji}$ \\
        BDE Distribution Shift & $\Delta_{\text{BDE}}$ & Mean deviation from reference BDEs \\
        BDE Distribution Width & $\sigma_{\text{BDE}}$ & Standard deviation of $D_{ij}$ distributions \\
        2$D$ Map Information Entropy & $H_{2D}$ & Shannon entropy of the $D_{ij}$ vs. $A_{ij}$ map \\
        \bottomrule
    \end{tabular}
\end{table}

\section{Results}
\label{sec:results}

\subsection{Emergence of BDE via Decomposable Learning}\label{sec:results_bde_emergence}

\begin{figure}[htbp]
    \centering
    \includegraphics[width=0.8\linewidth]{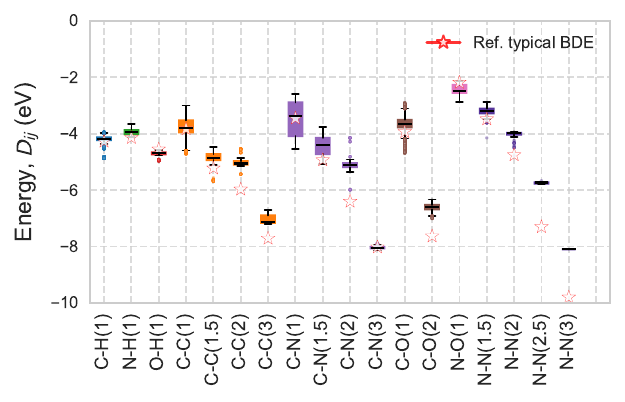}
    \caption{Emergence of BDE through E3D analysis for \ce{C-H}, \ce{N-H}, \ce{O-H}, \ce{C-C}, \ce{C-N}, \ce{C-O}, \ce{N-O}, and \ce{N-N} bonds. Distributions of learned symmetric bond energies ($D_{ij}$) from Allegro model (tensor size 64) trained on SPICE 2 dataset~\cite{Eastman-NutmegSPICELearning-2024y}, compared with reference BDE values (star markers)~\cite{Speight-LangesHandbookEdition-2016u}. The model successfully captures energy hierarchies and chemical trends without explicit bond training. The numbers (1), (2), and (3) represent single, double, and triplet bonds, respectively. The numbers (1.5) and (2.5) represent single and double bonds, respectively, in aromatic rings.}
    \label{fig:fig2}
\end{figure}

Our E3D analysis first reveals that Allegro models spontaneously develop physically meaningful bond energy via decomposable learning. This emergent understanding of bond energetics is crucial for accurately modeling reaction energies ($\Delta E$) and offers foundational insight before addressing the more complex scaling of activation energies ($E_a$). Figure~\ref{fig:fig2} shows distributions of internally derived symmetric bond energies ($D_{ij}$, Eq.~\ref{eq:DijAij}) for common chemical bonds from a SPICE~2 trained model, compared with reference BDE values.

The model yields $D_{ij}$ distributions that align remarkably well with reference BDE trends across diverse s. First, for \ce{C-H}, \ce{N-H}, and \ce{O-H} bonds, their strengths are correctly ordered, and their average energies quantitatively agree well with the reference. More significantly, the model correctly captures fundamental chemical trends for \ce{C-C}, \ce{C-N}, \ce{C-O} and \ce{N-N} bonds, successfully distinguishing their single, double, and triple bond variations with energy differences consistent with chemical intuition.
Each  exhibits a distribution spread of approximately 0.1 to 1.0 eV, likely reflecting environmental dependencies. This phenomenon is characterized herein as the 'emergence of BDE'.

This performance is achieved by learning only from total system energies and forces. It demonstrates the model's acquisition of fundamental bonding principles via decomposable learning.

\subsection{Robustness Across Training Domains}\label{sec:results_robustness_matpes}

To test robustness of the emergent BDE, we trained Allegro models exclusively on MatPES dataset\cite{Kaplan-FoundationalPotentialMaterials-2025p}. This dataset primarily consists of non-molecular inorganic structures lacking explicit molecular bonds. Even these MatPES-trained models, when evaluated on molecular test structures,  yield $D_{ij}$ distributions qualitatively agree with SPICE-trained models and aligned with reference BDE trends (visualized in Supplementary Information, Figure~\ref{fig:s1_matpes_bde}).

This emergence of BDE from bulk materials data indicates that the model learns a general, transferable understanding of local covalent interactions based on atomic environments, not specific molecular topologies. Thus, the trained model captures fundamental chemical bonding intuitions beyond particular training data domains.

\subsection{Internal Representation Refinement with Data Scaling}\label{sec:results_scaling_internal_rep}

Our E3D framework also enables detailed analysis of how the quality of internal representation evolves with dataset size. This analysis offers direct insights into the factors potentially contributing to the ``$E_a$ scaling wall.'' 

\begin{figure}[htbp]
    \centering
    \includegraphics[width=0.8\linewidth]{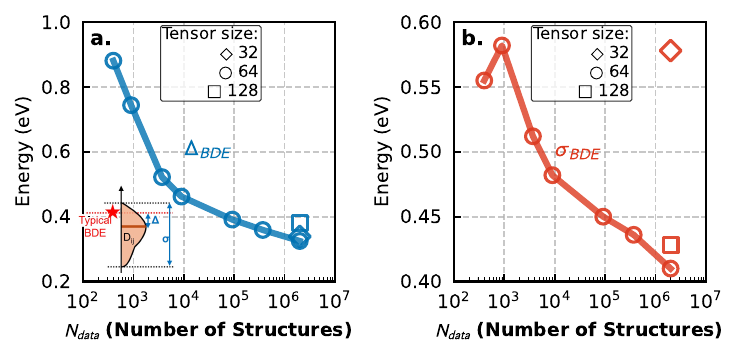}
    \caption{Evolution of BDE metrics with training data size from E3D analysis. (a) BDE distribution shift ($\Delta_{\text{BDE}}$): MAE between learned $D_{ij}$ means and reference BDEs. (b) BDE distribution width ($\sigma_{\text{BDE}}$): three standard deviations of learned $D_{ij}$ distributions. While $\Delta_{\text{BDE}}$ exceeded $10^4$, the slope became smaller and showed a tendency toward saturation. $\sigma_{\text{BDE}}$ continued to show a monotonic decreasing trend.}
    \label{fig:fig3}
\end{figure}

Figure~\ref{fig:fig3} highlights two key trends. First, the mean alignment with reference BDEs tends to saturate with a decreasing slope beyond $10^6$ training structures. Second, the distribution width continues to decrease, indicating increased confidence in the model prediction.

Our analysis has also revealed distinct influences of model size to $\Delta_{\text{BDE}}$ and $\sigma_{\text{BDE}}$ with respect to the number of structures used in the training. $\Delta_{\text{BDE}}$ is less sensitive to model size and shows a similar trend across the tensor sizes of 32, 64, and 128.  $\sigma_{\text{BDE}}$ on the other hand, is substantially influenced by the size of the model, where the larger tensor sizes of 64 and 128 show a continuous decreasing trend, suggesting increasing confidence in the prediction of the BDE. 

Taken together, our E3D analysis reveals that accurate and confident internal representations require both substantial training data and well-designed model capacity.

\begin{figure}[htbp]
    \centering
    \includegraphics[width=0.9\linewidth]{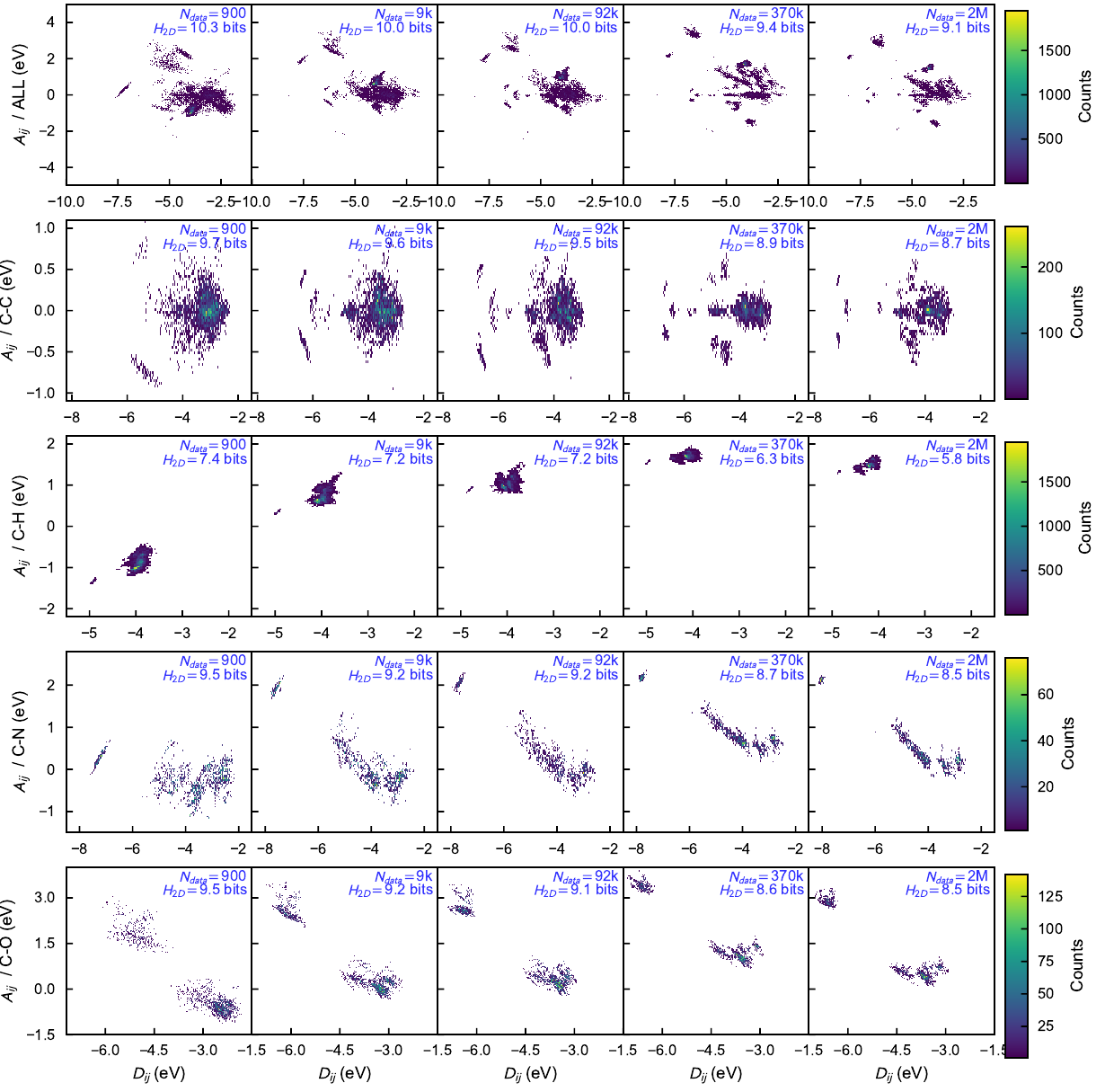}
    \caption{Evolution of learned energy representations through E3D analysis. 2$D$ histograms show symmetric ($D_{ij}$) vs. asymmetric ($A_{ij}$) bond energy distributions for \ce{C-H}, \ce{C-C}, \ce{C-O}, and \ce{C-N} bonds at varying training data volumes $N_{\rm data}$. Analysis performed on a consistent T1x reactant structures. Increased data leads to sharper, more structured representations, which is reflected in less Shannon entropy $H_{2D}$.}
    \label{fig:fig4}
\end{figure}

Beyond scalar metrics, our E3D framework also allows structural analysis using 2$D$ maps of symmetric ($D_{ij}$) versus asymmetric ($A_{ij}$) bond energy components. Figure~\ref{fig:fig4} shows these maps for overall bonding characteristics as well as for key chemical bond types such as \ce{C-H}, \ce{C-C}, \ce{C-O}, and \ce{C-N}, across various training conditions while increasing amounts of SPICE~2 data. The analysis was performed on the structures from T1x datasets to evaluate the model's generalizability to unseen data. 

A clear trend emerges: with expanding training data volume, the distributions for each bond type localize into distinct regions with sharper and clearer boundarys. This sharpening of distribution boundary and localization signify the development of more refined and structured internal energy representations by the training model.

Furthermore, as the model learns different bond types, these 'signatures' manifest as island-like distribution shapes in the 2$D$ maps. These patterns have been shown to be unique for each bond types, akin to fingerprints, which may be used as a potential indicator to contrast the differences in training progression among various bond types.

\subsection{Quantitative Analysis via Information Entropy}\label{sec:results_internal_rep_evolution}

\begin{figure}[htbp]
    \centering
    \includegraphics[width=1.0\linewidth]{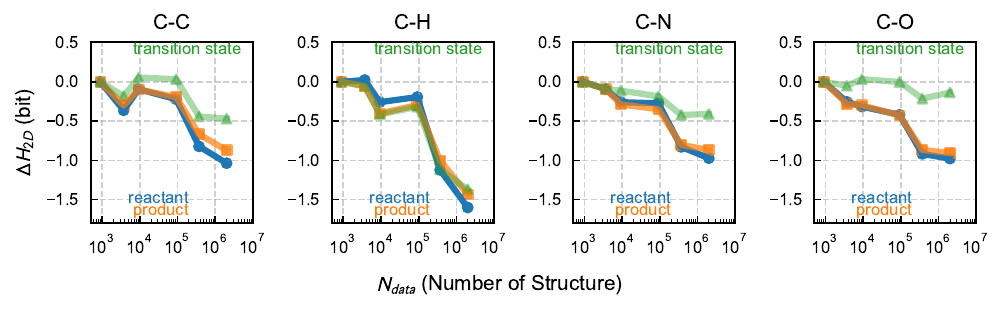}
    \caption{Quantitative evolution of internal representations with respect to the training data size. Shannon entropy ($H_{2D}$) of $D_{ij}$ vs. $A_{ij}$ maps decreases by each bond types. Here, $\Delta H_{2D}$ is shown relative to the entropy-based value at $N_{\rm data}$ = 900. Each line indicates T1x structures as reactants (blue), products (orange) and transition states (green), respectively.}
    \label{fig:fig5}
\end{figure}

The $D_{ij}$-$A_{ij}$ maps provide a qualitative visual evaluation of specific bond types. To quantify the structural organization observed in these maps, we employ Shannon entropy ($H_{2D}$) as a measure of the representational order.
Figure~\ref{fig:fig5} reveals a systematic decrease in entropy for bonds in the T1x dataset with varying SPICE~2 training data, confirming that internal representations become more organized.

Notably, a significant, sharp decrease in entropy occurred around $10^5$ data size. This suggests a transition point in learning where the model establishes clearer, more reliable internal representations of various bond types and their local environments. This transition resembles an "emergent ability" phenomenon\cite{Wei-EmergentAbilitiesModels-2022b,Berti-EmergentAbilitiesSurvey-2025s}, where new qualitative capabilities arise at specific data thresholds along with the previously reported learning scalability.

Comparison of the near-ground-state stable structures of reactants and products to the out-of-distribution TS structures has revealed distinct learning patterns across different bond types. While \ce{C-C} and \ce{C-H} bonds show similar entropy reduction trends in both stable and TS structures, \ce{C-N} and \ce{C-O} bonds exhibit markedly weaker entropy decreases specifically in TS configurations. This indicates that the model struggles to develop well-organized representations specifically for these bond types in transition states.

Significantly, this representational limitation coincides with the data range where the activation energy ($E_a$) scaling wall was observed (see Figure~\ref{fig:fig1} and \ce{C-N} and \ce{C-O} bonds in Figure~\ref{fig:fig5}). Our analysis identifies the root cause of the scaling wall and provides a valuable insight on how to overcome it.

\section{Discussion}\label{sec:discussion}

\subsection{Emergence of BDE phenomenon via Decomposable Learning:}\label{sec:discussion_emergence_decomposability}

Our E3D analysis demonstrates that a well-trained MLIP develops internal energy representations ($D_{ij}$) that correlate strongly with BDE (Figure~\ref{fig:fig2}). Remarkably, this occurs despite the model being trained only on total energy data—a phenomenon we term "emergence of BDE".

This emergent BDE capability improves systematically with increasing training data (Figure~\ref{fig:fig3}) and correlates with enhanced performance on reaction energies ($\Delta E$) (Figure~\ref{fig:fig4}). It also appears as an improved inference resolution as well as a reduction in information entropy, $H_{2D}$ (Figure~\ref{fig:fig5}). Furthermore, this capability proves robust across material data domains: models trained exclusively on bulk materials data (MatPES) develop transferable chemical principles applicable to molecular systems (Figure \ref{fig:s1_matpes_bde}). Our E3D framework for the first time has revealed that physically meaningful representations of chemical bonds and quantitatively comparable BDE values emerge without explicit supervision during MLIP trainig.

The development of classical interatomic potentials relies on prescribed functions and parameters. For example, in Morse potential, the bottom of the potential energy well directly corresponds to the BDE. Embedded Atom Method (EAM) for metals, as a example of many-body interatomic potentials, incorporates a distinctive many-body term through an embedding function that describes how the total cohesive energy is partitioned into a set of nonoverlapping one-body atomic energies~\cite{Daw-Embedded-atomMethodMetals-1984p}. 

In contrast, an MLIP autonomously learns how to decompose system-wide information (e.g. the cohesive energy) into physically meaningful sub-components, such as emergent BDEs, as the amount of training data increases. This emergent ability resembles phenomena observed in large language models~\cite{Wei-EmergentAbilitiesModels-2022b, Berti-EmergentAbilitiesSurvey-2025s}, suggesting that the emergence may be a general property of sufficiently large, well-trained neural networks applied to complex systems. While the mechanisms may differ, both systems exhibit qualitatively new capabilities that emerge from quantitative scaling.

Our results show that quantitative scaling in data and model parameters can lead to qualitatively new physical insights that emerge spontaneously rather than being explicitly programmed.

\subsection{Contrast with Direct BDE Training Approaches}\label{sec:direct BDE}
The emergent BDE capability suggests a novel paradigm contrast to the conventional approaches where a model is trained on explicitly defined target BDEs. Recent GNN-based models achieve accurate BDE predictions within 2 kcal/mol using curated datasets of thousands to tens of thousands of compounds~\cite{Sowndarya-ExpansionBondSpace-2023k,Gelzinyte-TransferableMachineMolecules-2024n}. This approach offers clear advantages in training efficiency, including small training datasets, rapid convergence, and straightforward expansion to new bond types through targeted libraries.

In contrast, our emergent BDE capabilities develop naturally from total energy and force training alone—without any explicit BDE information. While this emergent BDE requires larger and more diverse training datasets, a well-trained MLIP may acquire broader transferability by capturing fundamental chemical intuitions rather than be specialized for a specific set of BDEs.

This presents a critical trade-off in computational chemistry: targeted efficiency versus emergent generality. Direct BDE training provides immediate, quantitative results with minimal data requirements. The emergence of BDE offers the potential for broader chemical understanding at the cost of computational resources and training complexity.
The choice between these paradigms will likely depend on specific application requirements: whether users prioritize rapid deployment with known performance bounds, or seek models with potentially superior transferability across diverse chemical spaces.

\subsection{Overcoming "scaling wall" by Data Diversity}\label{sec:results_Hybrid_dataset}

Information entropy derived from E3D analysis (Figure~\ref{fig:fig5}) has been shown to provide a clue for overcoming the $E_a$ scaling wall (Figure~\ref{fig:fig1}). Building on this, we incorporate data diversity—a concept explored in previous studies\cite{Nomura-Allegro-FMTowardsSimulations-2025u,Tomoya-TamingMulti-domainSimulations-2024e}—as an additional strategy. As shown in Table~\ref{tab:tabs_hybrid_acc}, training with a hybrid dataset that combines SPICE~2 and MatPES data improves activation energy prediction, achieving an $E_a$ MAE of 0.44~eV compared to 0.58~eV for models trained only on SPICE.

\begin{table}[htbp]
    \caption{Performance improvements with hybrid dataset training on T1x reactive properties.}
    \label{tab:tabs_hybrid_acc}
    \centering
    \begin{tabular}{lccc}
        \toprule
        Dataset & Tensor size & $E_a$ MAE (eV) & $\Delta E$ MAE (eV) \\ 
        \midrule
        \multirow{2}{*}{SPICE 2} & 64 & 0.61 & 0.31\\ 
        & 128 & 0.58 & 0.33\\
        \midrule
        \multirow{2}{*}{Hybrid} & 64 & 0.49 & 0.27\\
        & 128 & \textbf{0.44} & \textbf{0.25}\\ 
        \bottomrule
    \end{tabular}
\end{table}

The hybrid dataset improves prediction accuracy, because it includes structural information missing from the SPICE dataset alone. Although the MatPES data within the hybrid dataset doesn't include unstable structures like transition states, it contributes valuable inorganic solid-state bonding patterns. These patterns are quite different from the organic compounds found in SPICE. Learning from these distinct inorganic patterns allows the model to develop more generalized representations of atomic interactions. As a result, the $D_{ij}$ vs. $A_{ij}$ map for T1x transition states (shown in Figure~\ref{fig:s2_ts_maps}) changes shape significantly when the model is trained with the hybrid dataset compared to training with only SPICE. Notably, the map flattens considerably along $A_{ij}$ axis, especially for \ce{C-N} and \ce{C-O} bonds. We speculate that these fundamental changes in the map's shape may be one of the reasons for the improved $E_a$ performance.

Though the hybrid dataset has improved the predictability of $E_a$, we also observe a slight increase in ($H_{2D}$) of the $D_{ij}$-$A_{ij}$ maps, as shown in Figure~\ref{fig:s2_ts_maps}. A detailed comparison with models trained only on SPICE~2 (using the same tensor size) is given in Table~\ref{tab:entropy_comparison_table}.  
This increase in entropy with improved $E_a$ accuracy might seem counterintuitive, however, it could be an indicator that the trained model has acquired more diverse and richer representations from the hybrid dataset, otherwise not possible.

We note that to maximally take advantage of diverse, large datasets —and to simultaneously achieve highly structured (low-entropy) internal representations across these broad chemical spaces— a sufficient model expressivity is required. This could mean using larger tensor sizes or more sophisticated architectures, see Table~\ref{tab:tabs_hybrid_acc}.

\section{Outlook}\label{sec:outlook}

\subsection{E3D Framework as a Diagnostic Tool}
Our E3D framework, which analyzes distributions of symmetric ($D_{ij}$) and asymmetric ($A_{ij}$) bond energy components alongside their information entropy ($H_{2D}$), serves as a powerful diagnostic tool that extend beyond general error metrics. It offers both qualitative and quantitative insights into how MLIPs learn to represent specific local chemical environments and phenomena.

For instance, by visualizing the $D_{ij}$-$A_{ij}$ maps, E3D analysis allows a qualitative assessment of how well the model distinguishes different bond types, revealing the formation of distinct "energetic fingerprints" that sharpen with increased data (as seen in Figure~\ref{fig:fig4}). Quantitatively, the systematic reduction in $H_{2D}$ (Figure~\ref{fig:fig5}) confirms that the model develops more ordered and less ambiguous internal representations for stable structures as it learns. This can also highlight potential dataset limitations; for example, \ce{C-N} and \ce{C-O} bonds reach entropy plateau earlier than \ce{C-C} or \ce{C-H} bonds in SPICE~2-trained models, indicating fewer diverse environments for these heteroatomic bonds in the dataset.

\subsection{Training Strategies using E3D analysis} E3D analysis provides insights into exploitation and exploration of parameter space, which can facilitate the development of training strategies to enhance the MLIP performance on reactive tasks.

Firstly, as a method that directly utilizes $H_{2D}$ reduction, regularization of entropy-based loss during training can be considered. Entropy-based loss regularization is an established method in the field of deep learning. However, implementing entropy estimation under high-dimensional latent spaces poses significant challenges, and effective reduction of dimensionality remains an active area of research\cite{Meni-Entropy-basedGuidancePerformance-2024t}. The proposed entropy $H_{2D}$ is also calculated from 2$D$ features derived from multidimensional latent space, thus aligning with a similar approach.

Second, the analysis facilitates exploring diverse data using the $D_{ij}$-$A_{ij}$ maps, particularly when augmenting datasets for fine-tuning. For instance, when sourcing from large-scale reaction dataset~\cite{Chanussot-OpenCatalystChallenges-2020b,Schreiner-Transition1x-Potentials-2022b}, one can specifically select structures located in region of the map distant from established data clusters. To mitigate rapidly increasing development costs, research into efficient data curation for foundational models is a critical priority\cite{Kaplan-FoundationalPotentialMaterials-2025p,Schwalbe-Koda-InformationTheoryThermodynamics-2024z}.

\subsection{Broader Applications}
Metrics in E3D have many applications other than training, but here we will briefly introduce two examples of application in inference and novel opportunity utilize decomposable learning.
 
Firstly, the emergent BDE can serve as collective variables (CVs) in enhanced sampling simulations \cite{Noe-CollectiveVariablesMethods-2017s,Bochkarev-GraphAtomicPassing-2024l}. Employing BDE as a CV over the more conventional bond length represents a promising approach, as BDE is a direct measure of dissociation. Second, monitoring of $D_{ij}$ and $A_{ij}$ time evolution during molecular dynamics simulations visualizes the progress of the bond dissociation process. These tools provide a comprehensive framework to investigate a complicated reaction dynamics.

Finally, we mention the potential for extending decomposable learning beyond BDE. For example, dipole moment is promising for decomposable learning due to the existence of MLIPs that have been trained directly on this property\cite{Unke-PhysNetNeuralCharges-2019w,Gokcan-LearningMolecularNetworks-2022b}. Both emergent BDE and decomposed dipole moments can be interpreted as abstract, partial representations of local electron density, fundamentally derived from Density Functional Theory (DFT). We hope that physical properties currently accessible only through DFT calculations, beyond the capabilities of conventional interatomic potentials, will gradually become computable using foundation models that incorporate decomposable learning techniques.

\subsection{Limitations}
\label{sec:limitations}
Our analysis is primarily based on the Allegro architecture. The extent to which these findings generalize to other MLIP families (e.g., MACE~\cite{Batatia-MACEHigherFields-2022t, Kovacs-MACE-OFF23TransferableMolecules-2023f}, NequIP~\cite{Batzner-E3-equivariantGraphPotentials-2022r}) requires further investigation.

\section{Conclusion}\label{sec:conclusion}

This study analyzed the implicitly learned energy decomposition characteristics within the Allegro MLIP architecture to explore its generalization behavior. We introduced and applied our \textbf{Edge-wise Emergent Energy Decomposition (E3D) analysis framework} via \textbf{decomposable learning}. This framework uses symmetric ($D_{ij}$) and asymmetric ($A_{ij}$) energy components, derived from the model's internal representation, to move beyond standard accuracy metrics, such as MAE and RMSE values on test data.

Key findings from our E3D analysis are:
\begin{itemize}
    \item \textbf{Emergence of BDE:} We found that the Allegro model learns an internal decomposed energy ($D_{ij}$) that strongly correlates with chemical BDEs across various bond types, effectively capturing bond order hierarchy.
    \item \textbf{Robustness and Universality:} E3D analysis demonstrated the consistency of this emergent BDE, even with training exclusively on non-molecular bulk materials data (e.g., MatPES). This indicates the learning of transferable, fundamental chemical principles.
    \item \textbf{Evolution of 2$D$ map metric:} $D_{ij}$-$A_{ij}$ maps and their information entropy ($H_{2D}$) showed that more training data systematically refines the internal representation, giving trainability information on each chemical bond type.
\end{itemize}

Our E3D analysis framework provides crucial insights into MLIP generalization mechanisms and limitations, offering a path beyond brute-force scaling. This framework offers promising approaches to develop more efficient, interpretable, and reliable MLIPs for a wide range of application in chemical reactions and other materials science fields.

{
\small
\section*{}
\bibliographystyle{unsrt} 
\bibliography{paperpile}
}

\clearpage
\appendix

\setcounter{table}{0} 
\renewcommand{\thetable}{S\arabic{table}}
\setcounter{figure}{0} 
\renewcommand{\thefigure}{S\arabic{figure}}

\section{Appendices}

\subsection{Software/Hardware environment}\label{sec:ap_hyperparameter}
\paragraph{Code}
This work used PyTorch version 2.5.1, NequIP version 0.6.1 from \url{https://github.com/mir-group/nequip}, and Allegro version 0.3.0 from \url{https://github.com/mir-group/allegro}.

\paragraph{Hyperparameter}
The hyperparameters used for training the Allegro-FM model, based on information extracted from arXiv:2502.06073, are detailed below. 

\begin{itemize}
    \item \textbf{Model Architecture (tensor size: 64)}
        \begin{itemize}
            \item Two-body latent MLP: Dimensions [64, 128, 256], SiLU nonlinearity.
            \item Later latent MLP: Dimensions [256, 256, 256], SiLU nonlinearity.
            \item Embedding MLP: Linear projection.
            \item Final edge energy MLP: Single hidden layer (dimension 128), no nonlinearity.
        \end{itemize}
    \item \textbf{Initialization:} Uniform distribution with unit variance for MLPs.
    \item \textbf{Cutoff:} Radial cutoff of 5.2 \AA.
    \item \textbf{Optimizer:} Adam with learning rate of 1e-3.
    \item \textbf{Batch Size:} The batch size was varied between 8 and 128, adjusted according to the size of the training dataset.
    \item \textbf{Weight Decay:} Default weight decay was used (value of 0 for `torch.optim.Adam`).
    \item \textbf{Epochs:} Training was performed until convergence was reached, determined by the learning rate scheduler monitoring the validation loss. A minimum of 100 epochs was executed for all training runs, including those using the largest SPICE 2 dataset.
    \item \textbf{Loss Function:} Combined Root Mean Square Error (RMSE) of per-atom energy, forces, and stress.
    \item \textbf{Loss Weights:} Energy : Force : Stress = 8 : 1 : 1.
\end{itemize}

\subsection{Dataset}\label{sec:ap_dataset}

The molecule targeted dataset utilizes SPICE version 2.0 \cite{Eastman-NutmegSPICELearning-2024y}. This significantly expanded dataset focuses on drug-like small molecules, peptides, and solvated amino acids relevant for bio-molecular simulations, now containing approximately 2 million conformations across nearly 114,000 molecules. Notably, version 2.0 extends the chemical space to include Boron (B) and Silicon (Si) containing molecules, alongside improved sampling of non-covalent interactions, while retaining forces and multipole moments calculated at the $\omega$B97M-D3(BJ)/def2-TZVPPD level of theory. Crucially, both dataset types underwent an energy offset alignment pre-processing step to ensure direct energetic comparability despite their different origins and calculation levels. Further details regarding the composition provided in Table \ref{tab:tabs1}.

For broader generalization studies, particularly to assess the emergence of physical scales from non-molecular data, we also used data from MatPES~\cite{Kaplan-FoundationalPotentialMaterials-2025p}, specifically its r²SCAN functional calculations; MatPES also contributed to our mixed-dataset training.
For direct analysis of reactive properties and detailed bond characteristics within stable versus transition states, we further used structures from the Transition1x (T1x) dataset~\cite{Schreiner-Transition1x-Potentials-2022b}, for which energies were recomputed (as detailed below) to align with the SPICE level of theory. The T1x dataset was mainly used as a benchmark test set for evaluating activation barriers and reaction energies.

For T1x dataset structures (reactants, transition states (TS), and products), we recalculated energies and forces using Psi4~\cite{Smith-Psi41Chemistry-2020c} at the SPICE dataset's theory level ($\omega$B97M-D3(BJ)/def2-TZVPPD).

\begin{table}[htbp]
  \centering
  \caption{Training datasets, MPtrj, MatPES, OFF (SPICE 1), and SPICE 2 are datasets with structures generated using PBE, r²SCAN, and $\omega$B97M-D3(BJ) methods respectively. $N_{data}$ is the total number of structures.}
  \label{tab:tabs1}
  \begin{tabular}{ccccccc}
    \toprule
     Name & SPICE 2& OFF & MatPES& MPtrj& $N_{data}$\\
      & & (SPICE 1)&&&&\\
    \midrule
    Method&$\omega$B97M-D3(BJ)&$\leftarrow$&r²SCAN&PBE&--\\
    Num. of structure&2.0 M&1.0 M&0.4 M&1.6 M&--\\
    \midrule
    SPICE 2  &$\checkmark$&&&&2.0 M\\
    MatPES   &&&$\checkmark$&&0.4 M\\
    Hybrid   &$\checkmark$&&$\checkmark$&&2.4 M\\
    \midrule
    Allegro-FM\cite{Nomura-Allegro-FMTowardsSimulations-2025u} &&$\checkmark$&&$\checkmark$&2.6 M\\
    \bottomrule
  \end{tabular}
\end{table}

\subsection{Typing for BDE metric evaluation}\label{sec:ap_bde}
This section provides details on the BDE metrics evaluation and reference BDE sources.

\paragraph{Manual Correction of GAFF typing}
We inferred the multiplicity for each bond from the GAFF types of the two atoms forming it. Any unclear automated assignments were manually corrected, and we added 0.5 to the multiplicity of aromatic ring bonds for our analysis. We then used these bond multiplicities, along with the element types of the atoms, to categorize bonds for our "Type-Decomposition" in the E3D analysis. This categorization was particularly used when calculating and comparing BDE metrics like $\Delta_{\text{BDE}}$ and $\sigma_{\text{BDE}}$.

\paragraph{Reference of literature BDEs}
Table~\ref{tab:bdes} lists these typical BDEs for a variety of common chemical bond types, along with their standard bond orders and the primary sources for these energy values. These reference BDEs, primarily from experimental measurements with some established theoretical values where experiments are unavailable (details in \cite{Speight-LangesHandbookEdition-2016u, And-TheoreticalBondEvaluation-2001q}), are fundamental, long-recognized measures of chemical bond strength.

\begin{table}[htbp]
    \centering
    \caption{Literature BDEs referenced from Lange's Handbook\cite{Speight-LangesHandbookEdition-2016u}\textsuperscript{a} and theoretical calculations\cite{And-TheoreticalBondEvaluation-2001q}\textsuperscript{b}.}
    \label{tab:bdes}
    \begin{tabular}{cccc}
        \toprule
        Bond & Bond order & Energy (eV) & Ref. structure \\
        \midrule
        \ce{C-H} & 1 & 4.27\textsuperscript{a} & \ce{H-CH2CH3} \\
        \ce{N-H} & 1 & 4.14\textsuperscript{a} & \ce{H-(NCH3)_2} \\
        \ce{O-H} & 1 & 4.55\textsuperscript{a} & \ce{H-OCH3} \\
        \ce{C-C} & 1 & 3.83\textsuperscript{a} & \ce{H3C-CH3} \\
        & 1.5 & 5.22\textsuperscript{b} & benzene \\
        & 2 & 5.97\textsuperscript{b} & \ce{-(HC)=(CH)-} \\
        & 3 & 7.71\textsuperscript{b} & \ce{HC#CH} \\
        \ce{C-N} & 1 & 3.45\textsuperscript{a} & \ce{(CH3)-NH2} \\
        & 1.5 & 4.93\textsuperscript{a} & (interpolated) \\
        & 2 & 6.41\textsuperscript{a} & \ce{(CH2)=NH} \\
        & 3 & 8.02\textsuperscript{a} & \ce{C#N} \\
        \ce{C-O} & 1 & 3.93\textsuperscript{a} & \ce{HO-CH3} \\
        & 2 & 7.63\textsuperscript{a} & \ce{H2C=O} \\
        \ce{N-O} & 1 & 2.18\textsuperscript{a} & \ce{HO-NCH3} \\
        & 1.5 & 3.59\textsuperscript{a} & (interpolated)\\
        & 2 & 5.01\textsuperscript{a} & \ce{HN=O} \\
        \ce{N-N} & 1 & 2.22\textsuperscript{a} & \ce{H2N-NHC6H5} \\
        & 1.5 & 3.48\textsuperscript{a} & (interpolated)\\
        & 2 & 4.75\textsuperscript{a} & \ce{HN=NH} \\
        & 2.5 & 7.30\textsuperscript{a} & (interpolated)\\
        & 3 & 9.84\textsuperscript{a} & \ce{N#N} \\
        \bottomrule
    \end{tabular}
\end{table}

\subsection{Dataset size dependency}\label{sec:ap_dataset_size_dependency}
This section provides detailed numerical data supporting the discussions on data and model scaling effects presented in the main text (e.g., Figure~\ref{fig:fig1} and Figure~\ref{fig:fig3}). Table~\ref{tab:tabs2} tabulates the Mean Absolute Errors (MAEs) for activation energies ($E_a$) and reaction energies ($\Delta E$) obtained from models with different tensor sizes. These results correspond to training on various subsets of the SPICE~2 and MatPES datasets, as well as on the Hybrid dataset, illustrating the impact of training data volume and diversity on predictive accuracy.

\begin{table}[htbp]
  \caption{Mean Absolute Error for activation energy ($E_a$), reaction energy ($\Delta E$) on various datasets with different model tensor sizes and training set sizes.}
  \label{tab:tabs2}
  \centering
  \begin{tabular}{lccccc}
    \toprule
     Dataset & Tensor & Num. of & MAE($E_a$) & MAE($\Delta E$) \\
     & size & Structure & [eV] & [eV]\\
    \midrule
    SPICE 2&64&400& 1.61& 0.93 \\
    & 64&   900 & 1.45& 0.87 \\
    & 64& 3.7 k & 1.04& 0.69 \\
    & 64& 9.2 k & 0.73& 0.56 \\
    & 64&  92 k & 0.60& 0.47 \\
    & 64& 370 k & 0.56& 0.42 \\
    & 32&   2 M & 0.63& 0.55 \\
    & 64&   2 M & 0.61& 0.31 \\
    & 128&  2 M & 0.58& 0.33 \\
    \midrule
    MatPES& 64 &4 k & 2.60& 1.35 \\
    & 64  &   9 k & 2.16& 1.02 \\
    & 64  &  77 k & 1.38& 0.77 \\
    & 64  & 390 k & 0.89& 0.60 \\
    & 128 & 390 k & 0.88& 0.57 \\
    \midrule
    Hybrid & 64 & 2.4 M & 0.49& 0.27 \\
    & 128 & 2.4 M & \textbf{0.44} & \textbf{0.25} \\
    \midrule
    Allegro-FM\cite{Nomura-Allegro-FMTowardsSimulations-2025u} & 128  & 2.6 M & 0.55 & 0.30 \\ 
    \bottomrule
  \end{tabular}
\end{table}

\begin{figure}[htbp]
    \centering
    \includegraphics[width=0.8\linewidth]{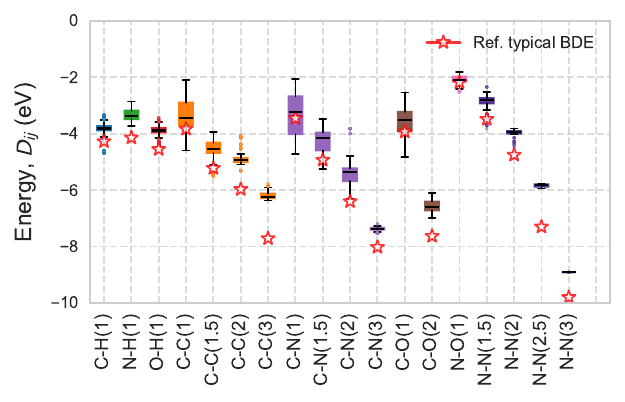} 
    \caption{Emergence of BDE through E3D analysis. Distributions of learned symmetric bond energies ($D_{ij}$) from an Allegro model (tensor size 128) trained on the MatPES dataset\cite{Kaplan-ScalingLawsModels-2020c}, compared with reference BDE values (stars markers). Distributions are qualitatively consistent with those from the SPICE~2-trained model (Figure~\ref{fig:fig2}).}
    \label{fig:s1_matpes_bde}
\end{figure}

\paragraph{Emergence of BDE on MatPES-Trained Model}
To further assess the robustness of the emergent BDE discussed in the main text (Section~\ref{sec:results_robustness_matpes}), this section presents E3D analysis results for an Allegro model trained exclusively on the MatPES dataset. Figure~\ref{fig:s1_matpes_bde} visualizes the learned symmetric bond energy ($D_{ij}$) distributions obtained from this model when evaluated on molecular structures. This analysis aims to demonstrate that the decomposable learning into physical extensive property is not restricted to training on molecular data alone, highlighting the model's ability to capture fundamental chemical principles from diverse data sources.

\begin{figure}[htbp]
    \centering
    \includegraphics[width=1.0\linewidth]{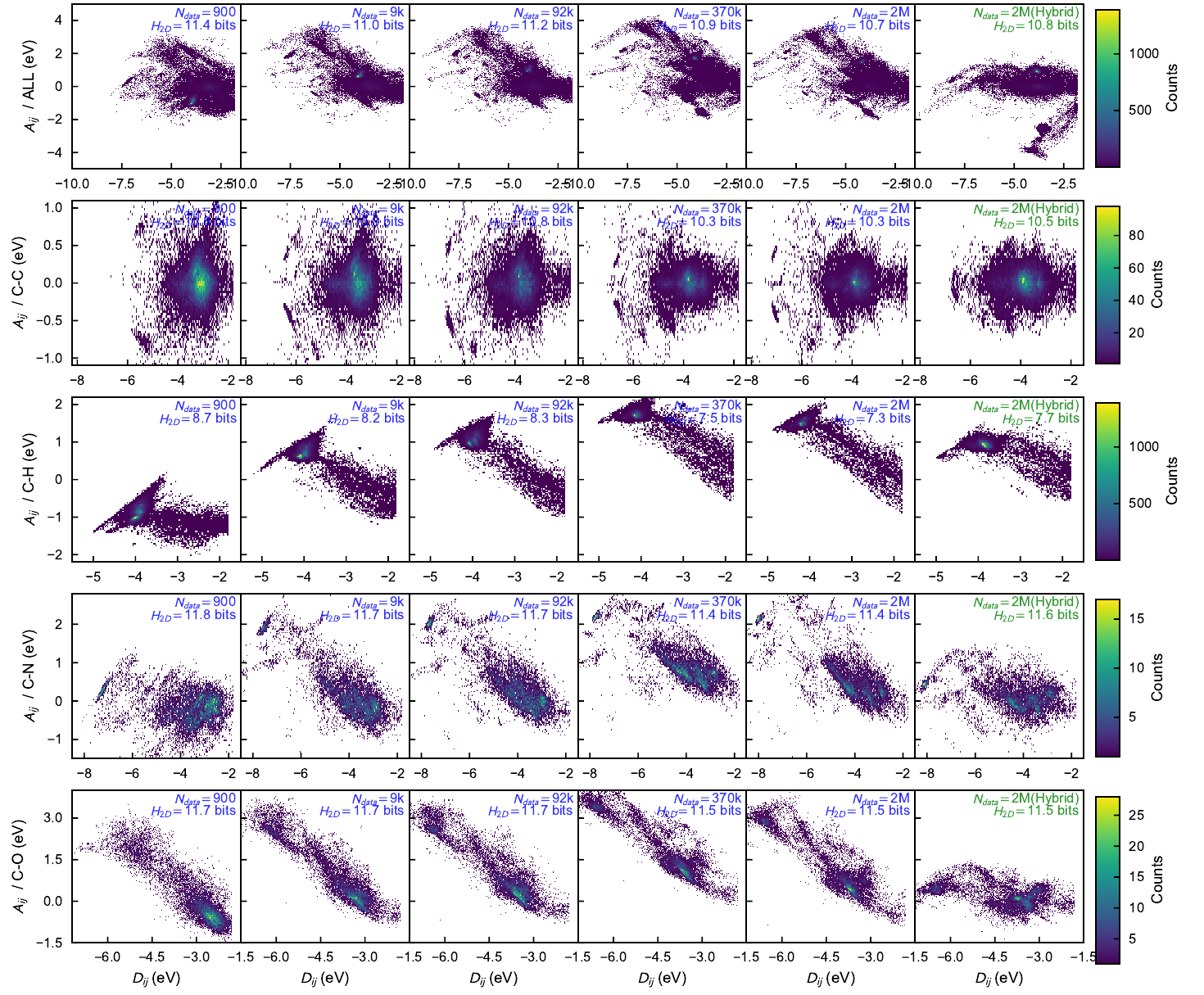} 
    \caption{Evolution of learned energy representations through E3D analysis. 2$D$ histograms show symmetric ($D_{ij}$) vs. asymmetric ($A_{ij}$) bond energy distributions for \ce{C-H}, \ce{C-C}, \ce{C-O}, and \ce{C-N} bonds at varying training data volumes $N_{\rm data}$. Analysis performed on a consistent T1x transition state structures.}
    \label{fig:s2_ts_maps}
\end{figure}

\paragraph{Internal Representations Refinement in Transition State Structures}
We plotted $D_{ij}$ vs. $A_{ij}$ distributions for transition state (TS) structures to compare with stable structures, as discussed in the main text (Figure~\ref{fig:fig4} in Sections~\ref{sec:results_internal_rep_evolution}). Figure~\ref{fig:s2_ts_maps} shows the more diffuse and less structured pattern for these high-energy TS configurations.

\subsection{Detailed analysis of information entropy and tensor size}
This section analyzes how model capacity (tensor size) and training dataset (SPICE 2 vs. Hybrid) affect the structure of learned internal representations using the E3D framework. We use Shannon entropy ($H_{2D}$) of the $D_{ij}$-$A_{ij}$ maps to measure this structure. Table~\ref{tab:entropy_comparison_table} summarizes these $H_{2D}$ values for different bond types in reactant and transition state structures, comparing models with 64 and 128 tensor sizes. The following paragraphs discuss key trends observed for each training condition.

\begin{table}[htbp]
\centering
\caption{Information entropy ($H_{2D}$) of $D_{ij}$ vs. $A_{ij}$ maps for models trained on SPICE~2 and Hybrid datasets with tensor sizes 64 and 128. Lower entropy indicates a more ordered internal representation.}
\label{tab:entropy_comparison_table} 
\begin{tabular}{@{}lllccccc@{}}
\toprule
Dataset & Structure & Tensor & \multicolumn{5}{c}{$H_{2D}$}(bits) \\
\cmidrule(lr){4-8}
        & Type      & Size   & ALL & \ce{C-C} & \ce{C-H} &\ce{C-O} & \ce{C-N} \\
\midrule
\multirow{4}{*}{SPICE} & \multirow{2}{*}{Reactant} & 64  & 9.06 & \textbf{8.65} & 5.82          & 8.51          & 8.52          \\
                    &                     & 128 & 9.04 & 8.92          & \textbf{5.58} & \textbf{8.4}  & \textbf{8.48} \\
\cmidrule(lr){2-8}
                    & \multirow{2}{*}{TS} & 64  & \textbf{10.66} & \textbf{10.28} & 7.28          & 11.53         & 11.43         \\
                    &                     & 128 & 10.71 & 10.43         & \textbf{7.15} & \textbf{11.4} & \textbf{11.27} \\
\midrule
\multirow{4}{*}{Hybrid}    & \multirow{2}{*}{Reactant} & 64  & 9.53 & 8.95          & 6.63          & 8.88          & 8.87          \\
                          &                     & 128 & \textbf{9.47} & \textbf{8.92} & \textbf{6.54} & \textbf{8.62} & \textbf{8.67} \\
\cmidrule(lr){2-8}
                          & \multirow{2}{*}{TS} & 64  & 10.83 & 10.51         & 7.7           & 11.55         & 11.65         \\
                          &                     & 128 & \textbf{10.7}  & \textbf{10.28} & \textbf{7.65} & \textbf{11.29} & \textbf{11.33} \\
\bottomrule
\end{tabular}
\end{table}

\paragraph{SPICE-Trained Models:}
Models trained only on SPICE showed different entropy pattern, when increasing the tensor size from 64 to 128, depending on the bond type.
For reactants, \ce{C-H} and \ce{C-O} bonds become more ordered with larger tensor size (\ce{C-H}: $5.82 \rightarrow 5.58$, \ce{C-O}: $8.51 \rightarrow 8.40$), showing that the more parameters refined representation of these bonds.
However, \ce{C-C} bonds in reactants showed slightly increased entropy($8.65 \rightarrow 8.92$), suggesting that for this more diverse bond type, larger model explored a broader representational space.
This shows that increasing model size doesn't automatically create better-ordered representations for all bond types, highlighting the balance between model capacity and chemical complexity.

\paragraph{Hybrid-Dataset-Trained Models:}
Training on the Hybrid dataset, which adds diverse elements and bonding environments from MatPES, revealed interesting patterns.

Hybrid models shows higher entropy values than SPICE models for the same organic bonds and tensor size, particularly for reactants (SPICE $9.06$ $\rightarrow$ Hybrid $9.53$ for tensor size 64).
This occurs because the model's representational capacity must cover more chemical elements and environments in the Hybrid dataset, creating less specialized representations for any specific bond type.

Importantly, increasing tensor size from 64 to 128 in Hybrid models consistently lowered entropy across all bond types in both reactants and TS (Reactant $9.53 \rightarrow 9.47$, TS  $10.83 \rightarrow 10.7$, for ALL bonds).
This decrease shows that the diverse Hybrid dataset requires at least 128 dimensions to properly represent its complex chemical environments.

\subsection{Cross-Domain Validation}
Beyond our main work on carbon-based bonds, we also explored how accurately our models predict energies for chemical systems that are uncommon in the training data, particularly those involving \ce{Si-O-H} bonds. We conducted these evaluations using general test sets split from our training data, which were different from the specific t1x reaction systems we focused on previously, as shown in Table~\ref{tab:tab_sioh_acc}. Usually, models trained only on a single type of dataset are expected to perform better when the test data is similar to their training data. Remarkably, however, our study showed that models trained with the hybrid dataset performed better in certain aspects—like predicting energies for both the MatPES and SPICE datasets and forces for the SPICE dataset—than models trained exclusively on either MatPES or SPICE data alone. This key finding strongly suggests that the hybrid dataset learned to generalize more effectively by being exposed to a wider range of chemical information. This improved generalization, in turn, allowed it to accurately understand and represent even infrequently seen chemical structures, such as \ce{Si-O-H} bonds.

\begin{table}[htbp]
    \caption{Cross-domain validation: RMSE performance on Si-O-H test subsets}
    \label{tab:tab_sioh_acc}
    \centering
    \begin{tabular}{cccccc}
        \toprule
         & \multicolumn{3}{c}{RMSE on MatPES} & \multicolumn{2}{c}{RMSE on SPICE 2} \\
        \cmidrule(lr){2-4} \cmidrule(lr){5-6}
        Trained Dataset & \textit{E} & \textit{F}$_{\ce{Si}}$ / \textit{F}$_{\ce{O}}$& \textit{P}& \textit{E}& \textit{F}$_{\ce{Si}}$ / \textit{F}$_{\ce{O}}$ \\
        &(meV/atom) &(meV/Å)&(meV/Å$^3$)&(meV/atom)&(meV/\AA) \\
        \midrule
        MatPES & 71 & \textbf{293 / 339} & 19 & 104 & 1080 / 327 \\
        SPICE 2 & 579 & 1326 / 1022 & 37 & 22 & 149 / 67 \\
        Hybrid & \textbf{50} & 585 / 394 & \textbf{13} & \textbf{6} & \textbf{107 / 59} \\ 
        \bottomrule
    \end{tabular} 
\end{table}

\end{document}